\documentclass[aps,pre,epsf,superscriptaddress,amsmath,amssymb,amsfonts,twocolumn,showpacs]{revtex4-1}

\usepackage{graphicx}
\usepackage{epsfig}
\usepackage{dcolumn}
\usepackage{bm}
\usepackage{braket}
\usepackage{amsmath}
\usepackage{color}

\begin{document}
\title{Many-Body Quantum Dynamics in the Decay of Bent Dark Solitons\\ of Bose-Einstein Condensates}

\author{G.C. Katsimiga}
\affiliation{Zentrum f\"{u}r Optische Quantentechnologien,
Universit\"{a}t Hamburg, Luruper Chaussee 149, 22761 Hamburg,
Germany}  

\author{S.I. Mistakidis}
\affiliation{Zentrum f\"{u}r Optische Quantentechnologien,
Universit\"{a}t Hamburg, Luruper Chaussee 149, 22761 Hamburg,
Germany}

\author{G.M. Koutentakis}
\affiliation{Zentrum f\"{u}r Optische Quantentechnologien,
Universit\"{a}t Hamburg, Luruper Chaussee 149, 22761 Hamburg,
Germany}\affiliation{The Hamburg Centre for Ultrafast Imaging,
Universit\"{a}t Hamburg, Luruper Chaussee 149, 22761 Hamburg,
Germany}
\author{P. G. Kevrekidis}
\affiliation{Department of Mathematics and Statistics, University
of Massachusetts Amherst, Amherst, MA 01003-4515, USA }
\author{P. Schmelcher}
\affiliation{Zentrum f\"{u}r Optische Quantentechnologien,
Universit\"{a}t Hamburg, Luruper Chaussee 149, 22761 Hamburg,
Germany} \affiliation{The Hamburg Centre for Ultrafast Imaging,
Universit\"{a}t Hamburg, Luruper Chaussee 149, 22761 Hamburg,
Germany}

\date{\today}

\begin{abstract}

The beyond mean-field dynamics of a bent dark soliton embedded in a two-dimensional repulsively interacting Bose-Einstein 
condensate is explored. 
We examine the case of a single bent dark soliton comparing the mean-field dynamics to a correlated approach, 
the Multi-Configuration Time-Dependent Hartree method for Bosons. 
Dynamical snaking of this bent structure is observed, signaling the onset of 
fragmentation which becomes significant during the vortex nucleation. In contrast to the mean-field approximation ``filling" 
of the vortex core is observed, leading in turn to the formation of
filled-core vortices, instead of the mean-field vortex-antivortex pairs. 
The resulting smearing effect in the density is a rather generic feature, occurring when solitonic structures are exposed to 
quantum fluctuations.  
Here, we show that this filling owes its existence to the dynamical building of an antidark structure developed 
in the next-to-leading order orbital. 
We further demonstrate that the aforementioned beyond mean-field dynamics can be experimentally detected using 
the variance of single shot measurements. 
Additionally, a variety of excitations including vortices, oblique 
dark solitons, and open ring dark soliton-like structures building upon higher-lying orbitals is observed. 
We demonstrate that signatures of the higher-lying orbital excitations emerge in the total density, and can be clearly 
captured by inspecting the one-body coherence. In the latter context, the localization of one-body correlations exposes the 
existence of the multi-orbital vortex-antidark structure.
\end{abstract}

\maketitle

\section{Introduction}
Bose-Einstein condensates (BECs) represent an ideal platform for the investigation of weak to strongly correlated 
quantum many-body (MB) systems, and are especially appealing due to their exquisite experimental control~\cite{pethick,stringari}. 
Various excitations  
can robustly emerge in BECs. 
Among these, dark solitons~\cite{emergent,frantz} and vortices~\cite{Fetter,siambook} constitute
paradigmatic examples. 
Dark solitons are persistent one-dimensional (1D) 
nolinear excitations characterized 
by a density notch and a $\pi$ phase shift. These waveforms have been 
experimentally realized~\cite{burger,becker,weller,weller2}
in bosonic systems, and theoretically as well as experimentally
studied in a
number of other physical settings including among others nonlinear optics~\cite{zakharov,drummond,yuridavies,drummond1} 
and superfluid Fermi gases~\cite{antezza,liao,scott,yefsah,ku1,ku2}. 
In repulsively interacting atomic BECs, and close to zero temperature mean-field (MF) dictates that 1D dark solitons 
exist as stable configurations being described by the Gross-Pitaevskii equation (GPE)~\cite{frantz}. Such a stability however, 
is altered in the presence of quantum fluctuations. This feature has triggered a new era of theoretical investigations
regarding the fate of the so-called quantum dark solitons when MB effects are taken into 
account~\cite{sachadark3,sachadark2,sachadark1,mishmash1,mishmash2,martin,sacha,sven,sacha17}. 
In most of the aforementioned cases, a filling of the dark soliton notch 
as a result of the depletion of the condensate 
has been reported, being, in turn, related to the quantum dispersion of the dark soliton's 
position~\cite{sachadark1,sacha}.

On the other hand, vortices can be thought of as the 
two-dimensional (2D) counterparts 
of dark solitons. As such, vortices are characterized also by a density depletion; at the same time, they
are topologically protected states possessing quantized circulation and in the 
lowest charge configuration (singly quantized vortices) a $2\pi$ phase winding. 
Vortices have been theoretically predicted and experimentally observed both in nonlinear 
optics~\cite{swartzlander,mcdonald,pomeau}
and in BECs; see for a representative sample the
experimental works of~\cite{Matthews,Madison,Abo-Shaeer,neely,middelkamp,donadello,wilson,kwon,samson}. 
In the latter context, the number of vortices nucleated strongly depends on the rotating/stirring frequency 
of the bosonic gas when compared to the trapping frequency. This availability
of rotation in BECs stirred numerous theoretical 
works that investigated vortex nucleation and interactions both at~\cite{butts,kavoulakis,linn,vorov} and beyond the MF 
approximation~\cite{fetter1,bloch,cooper,saarikoski,cremon,wells,imran1}. 
Weak~\cite{fetter1}, moderate~\cite{imran}, and rapid~\cite{bloch} rotating regimes have been explored, 
leading to the formation of one to few singly quantized vortices,
and progressively (with increased stirring) of regular vortex patterns 
forming canonical polygons and vortex lattices.   
For the MB 
treatment of these coherent structures, 
diagonalization techniques have been developed~\cite{imran,ahsan}
which, however, bear the limitation of tackling few 
boson systems. The MB effects on vortex formation in rotating BECs for larger bosonic ensembles has been considered very 
recently~\cite{kaspar,weiner} where modes of hidden vorticity not visible in the total density of the system, have 
been identified.

The connection between dark solitons and vortices has been experimentally established both in nonlinear 
optics~\cite{mamaev,tikhonenko} 
and in BECs~\cite{Denschlag,anderson,dutton}. In these works,
a direct observation of vortex nucleation via the so-called ``snaking" (transverse) instability
has been accomplished. 
The latter refers to the decay of dark solitons when embedded e.g. in a 2D geometry, alias dark soliton stripes (DSS), 
into vortex-antivortex pairs. A significant volume of theoretical studies
examined this dynamical instability~\cite{muryshev1,feder,carr,muryshev2,brand1,brazhnyi,Reichl2,cetoli,mateo,bulgac,lombardi}
and the conditions under which it can be suppressed~\cite{infrared,kamchatnov1,ma,Reichl}.

However, to the best of our knowledge, even though a series of theoretical works has been devoted in studying 
single component quantum dark solitons and vortices separately beyond the MF approximation,
no such efforts exist regarding vortex nucleation as a result of the ``snaking" of dark solitons. 
In the present contribution the dynamical nucleation of vortices stemming from the decay of an already bent dark soliton (BDS) 
is investigated~\cite{Mironov}. 
Starting from such a bent state the benefit is twofold. 
The decay process is accelerated in a controllable manner, e.g. via a stronger bending. 
Moreover, given the nature of the initial condition and its predominant
snaking in regions of large curvature,
the region where vortices are going to nucleate is 
a-priori ``designated''. 
This way, we study the dynamical deformation of the so-called BDS both at and beyond the MF approximation. 
To take into account quantum fluctuations in the BDS dynamics, we
use the Multi-Layer Multi-Configuration Time-Dependent Hartree Method for bosons (ML-MCTDHB)~\cite{moulos,moulosx}
designed for simulating the quantum dynamics of bosonic ensembles in higher dimensions.
In particular, a systematic comparison of the MF approximation, where a single orbital captures the BDS dynamics, 
with the full MB (multi-orbital) soliton dynamics is considered.

It is observed that when the snaking of the BDS occurs signals the onset 
of fragmentation within the correlated approach. The progressive development of this dynamical deformation leads  
to the vortex nucleation.  
During this process, fragmentation becomes significant, a result that is directly captured 
by the behaviour of the variance of single shot measurements.   
The latter exhibits an increasing tendency during the evolution in sharp contrast to the MF approximation where it remains almost constant.  
A number of vortex-antivortex pairs is formed (this number is two for our particular case examples, being controlled
by the background density and trap strength), emerging  at the core and at the edges of the bosonic cloud.
Most importantly a filling of the above-mentioned vortex dipoles is observed, leading to the formation of ``filled core'' vortices, i.e. not 
fully dipped as the ones predicted by the MF approximation (for which
the density vanishes). The latter observation constitutes one of our central 
results being also compatible with earlier findings regarding 1D quantum dark solitons~\cite{sachadark3,sachadark2}. 
More importantly, we show that these filled core vortices can be experimentally detected by averaging several single 
shot images using high optical resolution, i.e. of the order of the healing length, being attainable by contemporary experimental methods \cite{qmicroscope1,qmicroscope2}.  
We demonstrate that this filling mechanism results from the emergence of an antidark structure, i.e. a density hump on top of 
the BEC background, building upon the next-to-leading order orbital. This way, a composite multi-orbital vortex-antidark 
structure, stemming from the interplay of the first two significantly populated natural orbitals, emerges in the MB
density. Furthermore, a variety of excitations including vortices, oblique dark solitons~\cite{el,amo}, i.e. elongated BDS structures 
with vortices or vortex paths at their edges, and open ring dark soliton-like structures~\cite{anderson} developing in 
higher-lying orbitals is observed. We showcase the presence of both the 
antidark solitons as well as the vortices formed in higher orbitals as localized and incoherent regions, 
respectively, in the one-body coherence function. Finally,
the emergence of 
interparticle correlations in the BDS dynamics is shown by inspecting the Von-Neumann entropy on the one- and two-body level.   

The presentation of our work is structured as follows. In Section II a theoretical background both at
the MF and the MB level is provided as well as the initial ansatz used to simulate the BDS dynamics.
Section III contains our numerical findings regarding the dynamical deformation of the BDS   
and the spontaneous vortex nucleation both in the single orbital MF case
and that of the MB correlated approach. In Section IV we summarize our findings and discuss future 
challenges. Appendix A briefly comments on our computational methodology, and delineates the convergence of our results.
Finally, in Appendix B we discuss the initial state preparation, i.e. the way that the BDS is embedded into ML-MCTDHB.

\section{Setup and solitonic ansatz}

DSS are nonlinear excitations observed in 2D repulsively interacting BECs.
Such excitations are characterized by a density depletion of the 2D BEC being either 
a straight or curved stripe soliton.  
In the latter case, we refer to them as BDSs and in the present work they will be the main focus.  
Within the MF approximation the 2D model where such states can be found
to arise, is the 2D GPE, being a variant of the nonlinear Schr{\"o}dinger
equation \cite{pethick,Gross,Dalfovo} with cubic nonlinearity that also
typically considers them in the presence of an external trap.  
A generic BDS is characterized by its position being generally parametrized by a chosen path $x(y)$, its inverse width
$d=1/\xi$, and by the so-called soliton's phase angle $a(y)$. 
In the above expressions, $\xi=1/\sqrt{g |\tilde{\phi}_0 (0,0)|^2}= 1.26$ is the healing length being inversely 
proportional to the background density $|\tilde{\phi}_0 (0,0)|^2$, while $g$ denotes the interparticle interaction.     
Additionally, the soliton's phase angle is associated with its
velocity $u(y)/c = \sin a(y)$, with $c=\sqrt{g n/m}$ being the speed of sound. 
Here, $n$ refers to the local particle density, and $m$ 
is the particle mass. 

In the following we consider the out-of-equilibrium dynamics of a BDS being initially at 
rest, i.e. $u(y)=0$ ($a(y)=\pi/2$), and embedded in the background
density $\tilde\phi_0(x,y)$. 
The wavefunction of the BEC reads \cite{Mironov} 
\begin{equation}
\begin{split}
\tilde \phi(x,y;t)\equiv \tilde\phi_0(x,y) \Big\{ &\cos a(y)~ \tanh \left[ d \left( x-x(y;t) \right)\right] \\
&+ i \sin a(y) \Big\}. \label{Eq:SDsoli}
\end{split}
\end{equation} 
The ``bending'' is initially introduced by $x(y;t=0)=-X_0 \cos\left(\frac{2 \pi y}{\ell_y}\right)$,
where $X_0$ refers to the modulation amplitude and $\ell_y$ is the modulation length. The resulting dynamics of $x(y,t)$ even 
in the MF level is still a subject of active investigation; see, e.g.,~\cite{aipaper}. 
Note that for $X_0=0$ the DSS forms a density dip along a line passing
through the center of the trap. 
The above expression represents an {\it approximate} initial profile of the
MF (i.e. single orbital) setting. The approximate nature of the profile
stems from the effective multiplication with the equilibrium background
$\tilde\phi_0(x,y)$ at least in the case where $\tilde\phi_0(x,y)$ is not a constant.  
We remark that $\tilde\phi_0(x,y)$ within the Thomas-Fermi limit assumes the
approximate form $\tilde\phi_0(x,y)=\sqrt{\frac{1}{gN}[\mu-V(x,y)]}$. 
Here, $V(x,y)$ denotes the 2D external trapping potential, $\mu$ refers to the chemical potential of the 
background density, and $N$ is the total number of
atoms. It is also worth mentioning at this point that the BDS state of Eq.~(\ref{Eq:SDsoli})
is not a stationary state of the system even at the MF limit. However, by initializing the dynamics with  
this bent structure, the breakup dynamics can be studied in a controllable fashion since the snaking process is accelerated
for this curved structure. Further adding to this, we can also infer  the location of the vortices to be nucleated in the 
later stages of the dynamics, with the latter emerging around the region of maximum curvature of the initially ``engineered'' BDS
(see also our findings below). Finally, we also note that such a BDS state  
can be prepared experimentally using the standard phase imprinting method with the aid of a mask such as a spatial light 
modulator~\cite{becker,Denschlag,andersonmask}. In particular, using two
of the potential ``arms'' of the configuration utilized in~\cite{andersonmask}
could naturally give rise to the configuration considered herein. 

In addition to the above-mentioned approximation, the realm of the MF ansatz 
itself implies that the constituting particles of the BEC are uncorrelated.
Therefore, the total MB wavefunction within the MF approximation 
is expressed as a product of the MF wavefunctions
\begin{equation}
\Psi_{MF} ({\bf r}_1,\dots, {\bf r}_N;t) = \prod_{i=1}^{N} \phi({\bf r}_i;t). \label{Eq:MFansatz}
\end{equation} 
Here,  
${\bf r}_i=\left( x_i, y_i \right)$ labels the spatial coordinate of the atoms and  
$\phi({\bf r}_i;t)$ denotes the time-evolved wavefunction within the MF approximation.  
The equation of motion for the MF ansatz of Eq.~(\ref{Eq:MFansatz}) yields the well-studied  
2D GPE (see also below).  

However, within the MCTDHB approach~\cite{cederbaum1,cederbaum2} all particle correlations are systematically 
included. 
Indeed, the MB wavefunction $\Psi_{MB} ({\bf r}_1,\dots, {\bf r}_N;t)$ is constructed by permanents built upon 
$M$ distinct time-dependent 2D single particle functions (SPFs)
\begin{equation}
\begin{split}
&\Psi_{MB} ({\bf r}_1,\dots, {\bf r}_N;t)= \sum_{\substack{n_1,\dots,n_M\\ \sum n_i=N}} A_{(n_1,\dots,n_M)}(t)\times \\ 
&\sum_{i=1}^{N!} \mathcal{P}_i
 \left[ \prod_{j=1}^{n_1} \varphi_1({\bf r}_j;t) \cdots \prod_{j=1}^{n_M} \varphi_M({\bf r}_j;t) \right]. \label{Eq:MCansatz}
 \end{split}
\end{equation} 
In the above expression $\mathcal{P}$ denotes the permutation operator exchanging the particle configuration within the SPF 
$\varphi_i({\bf r};t)$, $i=1,2,...,M$, 
and $A_{(n_1,\dots,n_{M})}(t)$ correspond to the time-dependent expansion coefficients of a particular permanent.   
$N$ refers to the total particle number and $n_i(t)$ is the occupation number of the $i$-th SPF. 
Following the McLachlan time-dependent variational 
principle~\cite{McLachlan} for the generalized ansatz [see Eq.~(\ref{Eq:MCansatz})] yields the MCTDHB 
\cite{moulos,moulosx,cederbaum1,cederbaum2,matakias}  
equations of motion.  
These consist of a set of $\frac{(N+M-1)!}{N!(M-1)!}$ 
linear differential equations for the expansion coefficients and $M$ nonlinear integro-differential 
equations for the 2D SPFs $\varphi_i({\bf r};t)$.  
To the best of our knowledge analytical solutions of the MB ansatz that contain BDSs 
are not known, while systematic numerical studies in this direction are still lacking.  
Here, we utilize the MB variational approach that MCTDHB provides~\cite{mlmctdh} and embed at $t=0$ the MF wavefunction [see Eq.~(\ref{Eq:MFansatz})] 
within the MB ansatz [see Eq.~(\ref{Eq:MCansatz})]. To achieve the latter, we consider 
$A_{n_1=N}(0)=1$, $A_{n_1 \neq N}(0)=0$  [see Eqs. 
(\ref{Eq:SDsoli}), (\ref{Eq:MCansatz})] for the expansion coefficients and
$\varphi_1({\bf r};0)= \tilde \phi({\bf r};0)$  for the SPFs. 
The MF ground-state is used as the background density $\tilde\phi_0({\bf r})$.
Summarizing, we initialize the MB quantum dynamics 
employing the MF initial state, aiming to examine how the single-orbital population will spontaneously give rise to 
higher orbital dynamics. 

The natural orbitals, $\phi_i({\bf r};t)$, are defined as the eigenfunctions of the  
one-body density matrix \cite{Titulaer,Naraschewski} and are normalized to unity. 
The spectral representation of the one-body density matrix reads 
\begin{equation}
\label{eq:4} \rho^{(1)} ({\bf r},{\bf r}';t) =N \sum\limits_{i=1}^{M}
{{n_{i}}(t){\phi _{i}}({\bf r},t)} \phi _{i}^*({\bf r}',t), 
\end{equation}
where $M$ refers to the used number of orbitals and $n_i$ denotes the corresponding eigenvalues (natural populations).
Note here that for $M\rightarrow \infty$, $\rho^{(1)} ({\bf r},{\bf r}';t)$ tends to the exact one-body density 
$\tilde{\rho}^{(1)} ({\bf r},{\bf r}';t)$.  
In case our MB wavefunction $\Psi_{MB} ({\bf r}_1,\dots, {\bf r}_N;t)$ reduces to the
MF one [i.e. $\Psi_{MB} ({\bf r}_1,\dots, {\bf r}_N;t) \to \Psi_{MF} ({\bf r}_1,\dots, {\bf r}_N;t)$] 
the corresponding natural occupations obey $n_1(t)=1$, $n_{i\neq1}(t)=0$.    
In the latter case the first natural orbital $\phi_1({\bf r};t)$ reduces to the MF wavefunction $\phi({\bf r};t)$.  
Finally, we remark that the above-mentioned population eigenvalues $n_{i}(t) \in [0,1]$ characterize the so-called 
fragmentation of the system \cite{Penrose,Mueller}: For only one macroscopically occupied orbital the system is said to be 
condensed, otherwise it is fragmented. 

To examine the beyond MF dynamics of a BDS in a setting relevant to most of the recent experiments, 
we consider a bosonic gas trapped in a 2D harmonic oscillator potential. 
Such a ``pancake" geometry is experimentally realizable upon considering a strong confinement 
along the $z-$direction. 
The MB Hamiltonian consisting of $N$ bosons each with mass $m$ trapped in a 2D harmonic oscillator potential   
reads 
\begin{equation}
\begin{split}
H(\textbf{r}_1,\textbf{r}_2,...,\textbf{r}_N;t)= & \sum_{i=1}^{N} \left[ -\frac{\hbar^2}{2 m} \nabla^2_i 
+\frac{1}{2} m \omega_\textbf{r}^2 \textbf{r}_i^2 \right]\\ 
&+\sum_{i<j} V(\textbf{r}_i -\textbf{r}_j).\\
\end{split}
\label{Eq:SDsoli0}
\end{equation}
Here, $\omega_{\textbf{r}}=(\omega_x,\omega_y)$ where $\omega_x=\omega_y$
refers to the frequency of the isotropic external oscillator and $\textbf{r}_i=(x_i,y_i)$.  
The two-body $s$-wave interaction is modeled by a finite-range Gaussian shaped function~\cite{christensson,Doganov,Imran}
\begin{equation}
V(r_i-r_j)=\frac{g}{2\pi\sigma^2}e^{-\frac{(\textbf{r}_i-\textbf{r}_j)^2}{2\sigma^2}}, 
\label{Eq:int}
\end{equation}  
where $\sigma$ refers to the width of the Gaussian 
distribution. Note that the above interaction potential of  Eq.~(\ref{Eq:int}), tends to a contact interaction one as 
$\sigma\to 0$. In the following we consider only the dynamics of repulsively interacting bosons implying that $g>0$. 
In cold atomic gasses the interaction strength is related to the scattering length between the particles being experimentally 
tunable by Feshbach resonances~\cite{Inouye,Chin}. 
It has been shown \cite{Doganov} that in 2D and in the limit $\sigma/l_q\ll 1$ 
($l_q=\sqrt{\hbar/m \omega_q}$ refers to the harmonic oscillator 
length in the $q=x,y$ direction)  
the $s$-wave scattering length is related to the parameters of the Gaussian as $a_{2D}\approx\sqrt{2}\sigma e^{-\frac{\gamma}
{2}-\frac{\pi l^2}{g/\hbar \omega}}$ with  
$\gamma\approx0.577$ being the Euler-Mascheroni constant. 
For reasons of computational convenience, we shall set $\hbar=m=g=1$, and 
therefore all quantities below are given in dimensionless units ($g=1$
should be assumed everywhere unless otherwhise stated).
This way, the resulting dimensionless Hamiltonian [see also Eq.~(\ref{Eq:SDsoli0})] has two free parameters namely 
$\omega_x=\omega_y$ and $\sigma$. 
To examine the dynamical deformation of the BDS, the trapping frequency is fixed to $\omega_x=\omega_y=0.1$. 
Finally, we also choose $\sigma=0.2$ as a trade-off between smoothness and short range, when compared to the harmonic 
oscillator length, i.e. $\sigma<l_q$.

To initialize the beyond MF dynamics we first trace the MF ground state $\tilde \phi_0({\bf r})$, 
using a fixed point
(Newton-type) method. The latter is applied to the well-known GPE steady-state problem
\begin{equation}
\left[ -\frac{1}{2}\nabla^2 + \frac{1}{2} \omega_{\bf r}^2 {\bf r}^2+N | \tilde\phi_0 ({\bf r}) |^2 - \mu \right] \tilde\phi_0 ({\bf 
r}) =0.  
\end{equation}
On top of this relaxed MF background for a fixed number of atoms $N$, we then embed the BDS 
of Eq.~(\ref{Eq:SDsoli}) at $t=0$. 
We remark that by following the above-mentioned
procedure we minimize the sound wave emission during the dynamics.  
For more details on the 
selection of the soliton and background density parameters 
we refer the reader to Appendix \ref{sec:numerics}.

\section{Bent dark soliton dynamics}
\subsection{Comparing the mean-field and the many-body approach at the one-body density level}

Before delving into the MB BDS dynamics let us elaborate on the corresponding dynamical vortex nucleation at 
the MF level. 
It is worth mentioning at this point that such 2D bent structures are prone to decay \cite{Mironov}, leading  
to the formation of vortex-antivortex pairs, i.e., pairs of vortices having opposite circulation. The latter state has been 
argued to be quite robust at least at the MF level~\cite{middel10}. 
The above-mentioned decay of the BDS into vortex-antivortex pairs is observed
for all parameter values, namely for different modulation amplitudes,
i.e. $X_0=0.5, 0.8, 1.0$,
upon varying the trapping frequency within the interval $\omega_q=[0.002, 0.8]$, and also upon increasing the interaction 
strength, i.e. $g=[0.5, 2.0]$, that we have checked. 
We remark here, that by considering the BDS instead of a straight DSS we also break the parity symmetry 
along the $x$-direction while it is preserved along the $y$-direction. This parity symmetric direction further implies that 
indeed aligned vortices created perpendicular to this spatial direction,
as a result of the decay of the BDS, must have opposite circulations
(vortex-antivortex pairs).
Note also that throughout this work we fix the total number of atoms to $N=100$, while the modulation 
amplitude and modulation length are fixed to $X_0=1.0$, and $l_y=20$ respectively.
\begin{figure*}[ht]
\includegraphics[width=1.0\textwidth]{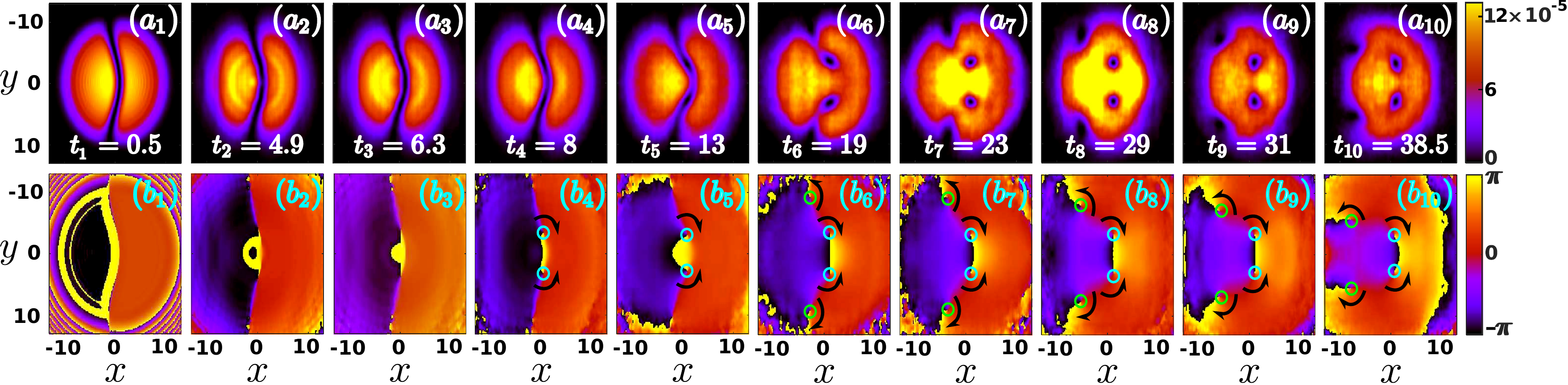}
\caption{(Color online) $(a_1)$-$(a_{10})$ Evolution of the density, $\rho^{(1)}(\textbf{r};t)$,
within the MF approximation at different propagation times (see legend).  
$(b_1)$-$(b_{10})$ show the corresponding phase, 
$\arg{[\phi(\textbf{r}_i;t)]}$, for the aforementioned  time instants.   
The initial BDS with $u=0$ possesses a modulation amplitude $X_0=1.0$ and period $l_y=20$, while the total number of bosons is 
$N=100$. The longitudinal and transverse confinement frequencies are $\omega_x=\omega_y=0.1$. Solid circles indicate the 
location, and arrows the circulation of the vortices nucleated.} \label{Fig:1}
\end{figure*}

\begin{figure*}[ht]
\includegraphics[width=0.98\textwidth]{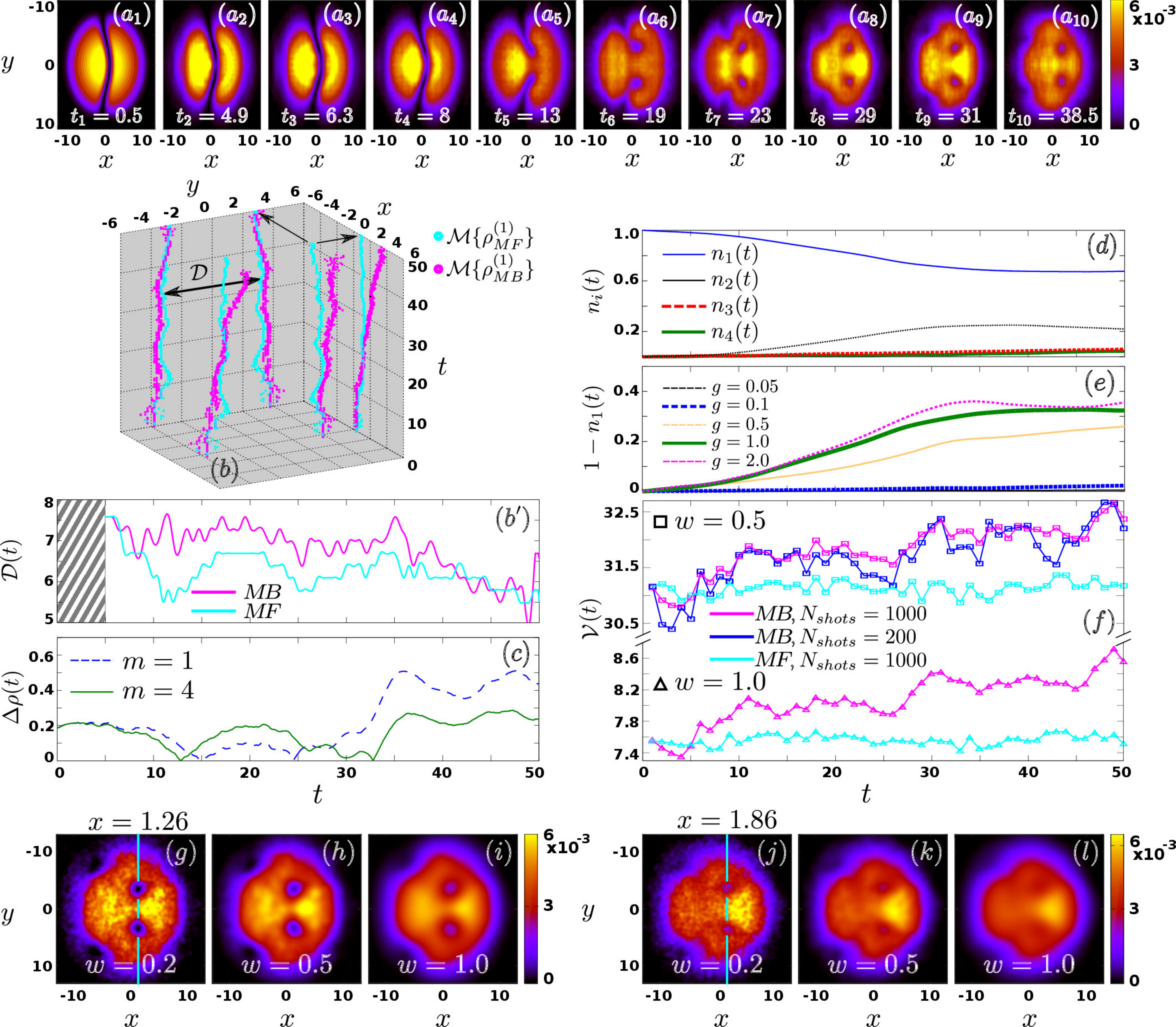}
\caption{(Color online) $(a_1)$-$(a_{10})$ One-body density, $\rho^{(1)}(\textbf{r};t)$, 
at different time instants during the evolution (see legend) calculated via MCTDHB.  
$(b)$ 3D plot showing the evolution of the density minima, $\mathcal{M}\lbrace\rho^{(1)}(t)\rbrace$, 
of the core vortex pair nucleated around the center of 
the trap, together with the relevant projections, indicated by black arrows, in the longitudinal $x$- and transverse $y$-
directions. Illustrated are the 
corresponding minima within the MF approximation and the MB approach (see legend). 
($b'$) Evolution of the inter-vortex distance $\mathcal{D}(t)$ on the MF and MB level (see legend). 
The dashed box indicates that for $t<5$ the vortices are not clearly formed and thus $\mathcal{D}(t)$ is not well defined. 
$(c)$ Evolution of the integrated total density imbalance, $\Delta \rho(t)$, within the MF i.e. $m=1$ approximation  
and the correlated $m=4$ approach. 
$(d)$ Natural populations $n_i(t)$, with $i=1,2,3,4$ of the four natural orbitals used within the correlated 
approach. $(e)$ Deviation from unity during propagation of the first natural occupation, $1-n_i(t)$, for different 
interaction strengths. ($f$) Evolution of the variance $\mathcal{V}$ obtained from in-situ single shot 
measurements within the MF approximation 
and the MB approach (see legend) for $g=1$. ($g$), ($h$), ($i$) [($j$), ($k$), ($l$)] Averaged images over 
$N_{shots}=1000$ at $t_9=31$ for different widths $w$ 
of the point spread function (see legend) within the MF approximation [MB approach]. 
Dashed blue lines in ($g$) and ($j$) denote the location in the longitudinal $x$-direction of the vortex dipole within the MF 
approximation and the MB approach respectively.  
Other parameters used are the same as in Fig.~\ref{Fig:1}.} 
\label{Fig:2}
\end{figure*}

In Figs.~\ref{Fig:1} $(a_1)-(a_{10})$, the total density, $\rho^{(1)}(\textbf{r};t)$, of the 
MF wavefunction is depicted for selected time instants up to $t_{max}=38.5$.  
Below each density, the corresponding phase, i.e. $\arg{\left[\phi(\textbf{r}_i;t)\right]}$,  
for the aforementioned instants is shown in Figs.~\ref{Fig:1} $(b_1)-(b_{10})$.   
It is found that from the very early stages of the dynamics, the snaking takes place.   
Notice the deformation that occurs at the core of the bent soliton shown in Fig.~\ref{Fig:1} $(a_2)$,
an event that is even more pronounced in its relevant phase illustrated in Fig.~\ref{Fig:1} $(b_2)$.  
As time evolves a dramatic change in the curvature of the BDS is observed,
and already at $t_4=8$ but more evidently at $t_5=13$ depicted in Fig.~\ref{Fig:1} $(a_5)$ a pair of vortices can be 
seen to start to form, with the two vortices being created
in the vicinity of the region of maximum curvature (core vortex pair). 
The relevant phase here, shown in Fig.~\ref{Fig:1} $(b_{5})$, provides a clearer picture  
since the $2\pi$ phase shift at the location of the formation of each of the aforementioned vortices, indicated by cyan 
circles, is evident. Following the trajectory of this vortex-antivortex pair
for larger propagation times, shown in Figs.~\ref{Fig:1} $(a_6)-(a_{10})$ 
it is observed that this quite robust vortex dipole remains trapped around the origin
$x=y=0$ with no further major change in the inter-vortex distance.   

Furthermore, at these later time instants another vortex pair is formed, being visible already in Fig.~\ref{Fig:1} $(a_6)$.
In contrast to the core vortex pair,  
this vortex dipole is created at the edges of the cloud (edge vortex pair) as is evident in the corresponding phase 
illustrated in Fig.~\ref{Fig:1} $(b_6)$ (see also the green circles indicating this new pair). 
Note that the location of the formation of this vortex dipole corresponds to the end points 
of the initially embedded BDS. It is observed that as time evolves this vortex pair
travels around the periphery of the cloud and towards the negative $x-$ direction, a motion that is followed by a decrease in 
the inter-vortex distance. In particular, this vortex pair is initially formed around $(x=-4, y=\pm 9)$ shown
in Fig.~\ref{Fig:1} $(a_6)$, while at later times depicted in Fig.~\ref{Fig:1} $(a_{10})$, it is located around 
$(x=-9, y=\pm 5)$.
However, at later times (results not shown here), this pair travels towards the center of the trap in an effort to penetrate 
the cloud with the relative distance between the two vortices remaining unchanged. Since it never gets trapped it reverses its 
motion travelling again towards the boundaries of the domain performing the above-mentioned epicyclic type of 
motion~\cite{middelkamp} for even larger propagation times. 

Having identified the dynamical decay of the BDS and the consequent vortex nucleation within the MF approximation, 
next we study the bent soliton dynamics, with the latter being initialized within the correlated multi-orbital 
approach (see also Appendix B).
For a direct comparison of the two different approaches the one-body density, $\rho^{(1)}(\textbf{r};t)$,
is illustrated in Fig.~\ref{Fig:2} $(a)$ for the same time instants as the ones shown in Fig.~\ref{Fig:1} for the MF case. 	
An overall qualitative agreement is observed between the MF and the 
MB approach, with the deformation of the BDS manifesting itself also within the correlated picture, and even around 
the same time scales. Notice the vortex pair formation around the center of the trap 
shown in Fig.~\ref{Fig:2} $(a_5)$  [see also Fig.~\ref{Fig:1} $(a_5)$], 
and also the second vortex pair created at the periphery of the cloud depicted e.g. in Fig.~\ref{Fig:2} $(a_7)$ 
[see here Fig.~\ref{Fig:1} $(a_7)$]. 
However, by the aforementioned comparison it also becomes apparent that while in the MF case 
both the core and the edge vortex dipoles
are ``fully'' dipped (i.e., the density vanishes at the vortex
cores), in the MB scenario {\it filled core} vortices are formed 
imprinting in this way even at the one-body density level their multi-orbital nature (since it is the additional orbitals that partially fill the core of
leading orbital vortex).

In an attempt to shed light on the differences observed between the two approaches, in 
Fig.~\ref{Fig:2} $(b)$ the trajectories of the density minima, $\mathcal{M}\lbrace\rho^{(1)}(t)\rbrace$, 
both at the MF and the MB
level are illustrated. 
To obtain this 3D plot, we calculated the minima at the core of the initially embedded BDS, 
for each of the above-mentioned densities. As the core of the BDS we identified the region around the center of the trap
within a radius $r=r_0=6$. We remark here, that within this radius we are not able to capture the dynamics 
around the endpoints of the BDS, and as a consequence the trajectory of the edge vortex pair. 
Notice, that at the very early stages of the dynamics, $t\lesssim 5$,
both approaches coincide.
Within the aforementioned time interval we present the minima that are proximal to the core vortex pair to be 
nucleated later on, instead of the entire line of minima that would correspond to the initial BDS. 
As time evolves and the nucleation of vortices as a result of the decay of the BDS takes place, i.e. at 
$t\approx 13$ or $t\approx 19$, the two approaches begin to deviate from one another. 
This deviation can be seen by inspecting the location of the calculated minima, corresponding from here on to the core of each 
of the two vortices that are nucleated aligned around the center of the trap. As it is observed, in the MB case these minima,
 $\mathcal{M}\lbrace\rho^{(1)}_{MB}(t)\rbrace$,
are found to be shifted towards slightly larger inter-vortex separation (that is along the $y$-direction) and also kicked 
further off of the center of the trap (along the $x$-direction), when compared to the MF approximation, i.e.  
$\mathcal{M}\lbrace\rho^{(1)}_{MF}(t)\rbrace$.
To quantify the shift along the $y$-direction in Fig.~\ref{Fig:2} $(b^{'})$ the evolution of the inter-
vortex distance 
$\mathcal{D}(t)$ is illustrated. Note that we measure $\mathcal{D}(t)$ after the vortex nucleation, i.e. for $t\geqslant 5$, 
both at and beyond the MF approximation. As is evident for times up to $t\approx40$ the vortices nucleated within the MB 
approach are slightly outer when compared to the MF ones, while for larger propagation times $\mathcal{D}(t)$ is almost 
the same for both approaches. 
On the other hand, the off-center kick of the core vortex pair becomes rather dramatic for larger 
propagation times, i.e. $t\gtrsim 25$,
a result that is evident in the projection along the $x$-direction in Fig. \ref{Fig:2} ($b$). This shift suggests an interplay between the leading 
order orbital and the higher-lying ones that we will trace in more detail later on. 

To further elaborate on the above-mentioned differences,   
Fig.~\ref{Fig:2} $(c)$ shows the evolution of the density imbalance both at the MF and the MB approach, being measured with 
respect to the origin and defined as $\Delta \rho (t)= \rho_L (t)-\rho_R (t)$. 
Here, $\rho_{L}(t)=\frac{1}{N}\int^{+\infty}_{-\infty}dy \int^0_{-\infty} dx\rho^{(1)}(x,y;t)$ 
$[ \rho_{R}(t)=\frac{1}{N} \int^{+\infty}_{-\infty}dy \int^{+\infty}_{0} dx\rho^{(1)}(x,y;t)]$ 
denotes the left [right] integrated 
density and $N=\int^{+\infty}_{-\infty}dy \int^{+\infty}_{-\infty} dx\rho^{(1)}(x,y;t)$. 
As expected, for the short time dynamics both approaches coincide. However, 
as time evolves the two approaches deviate from one another 
with the density imbalance being greater at the MF level 
most demonstrably at large propagation times
($30<t<50$) i.e. 
after the nucleation of vortices.
To gain further insight regarding the above-described quantitative difference between the two approaches,
next we study the evolution of the population of the natural orbitals, $n_i (t)$ with $i=1,\ldots,4$, depicted in  
Fig.~\ref{Fig:2} $(d)$. 
We remind the reader that a state with
$n_i (t) = 1$ is referred to as fully condensed, while for
$n_i (t)\neq 1$ the state is fragmented~\cite{Penrose,Mueller}. 
Since we initialize the dynamics at the MF level, $n_1(0)=1$ holds
and the first orbital is naturally expected to also dominate
the dynamics for the first time instants 
i.e. $n_1(0<t<2)\approx1$.
As time evolves, fragmentation is generally present 
being more pronounced in the first, $n_1(t)$, and second, $n_2(t)$, natural populations
and negligible for the higher-lying ones, namely $n_{3,4}(t)<0.1$.
The maximum slope, $\left(n_i(t)-n_i(t+\Delta t)\right)/\Delta t$, for $i=1,2$ 
occurs at intermediate times scales 
while it becomes nearly constant for larger propagation times.
This fragmentation rate can intuitively be connected with 
the density imbalance described
above as follows. In the absence of fragmentation (short time dynamics) the imbalance is larger for the MF case, while as fragmentation becomes 
significant the imbalance is greater in the MB approach ($13<t<25$)  
and finally when fragmentation tends to a constant value 
the measured imbalance becomes greater within the MF approach ($t>25$).     
Such a connection in turn suggests that $\Delta \rho(t)$ can be used to probe the fragmentation rate from the one-body 
density.
To elaborate on the interaction dependence of the fragmentation process the deviation from unity, $1-n_1(t)$, of the
first natural population is depicted in Fig.~\ref{Fig:2} $(e)$ for different interparticle repulsions. As it is observed near 
the non-interacting limit, i.e. $g=0.05,0.1$, 
fragmentation is highly suppressed
while as $g$ increases, i.e. $g=0.5,1,2$, the deviation from the MF approximation, imprinted in the presence of fragmentation,
becomes gradually more pronounced.
\begin{figure*}[ht]
\includegraphics[width=1.0\textwidth]{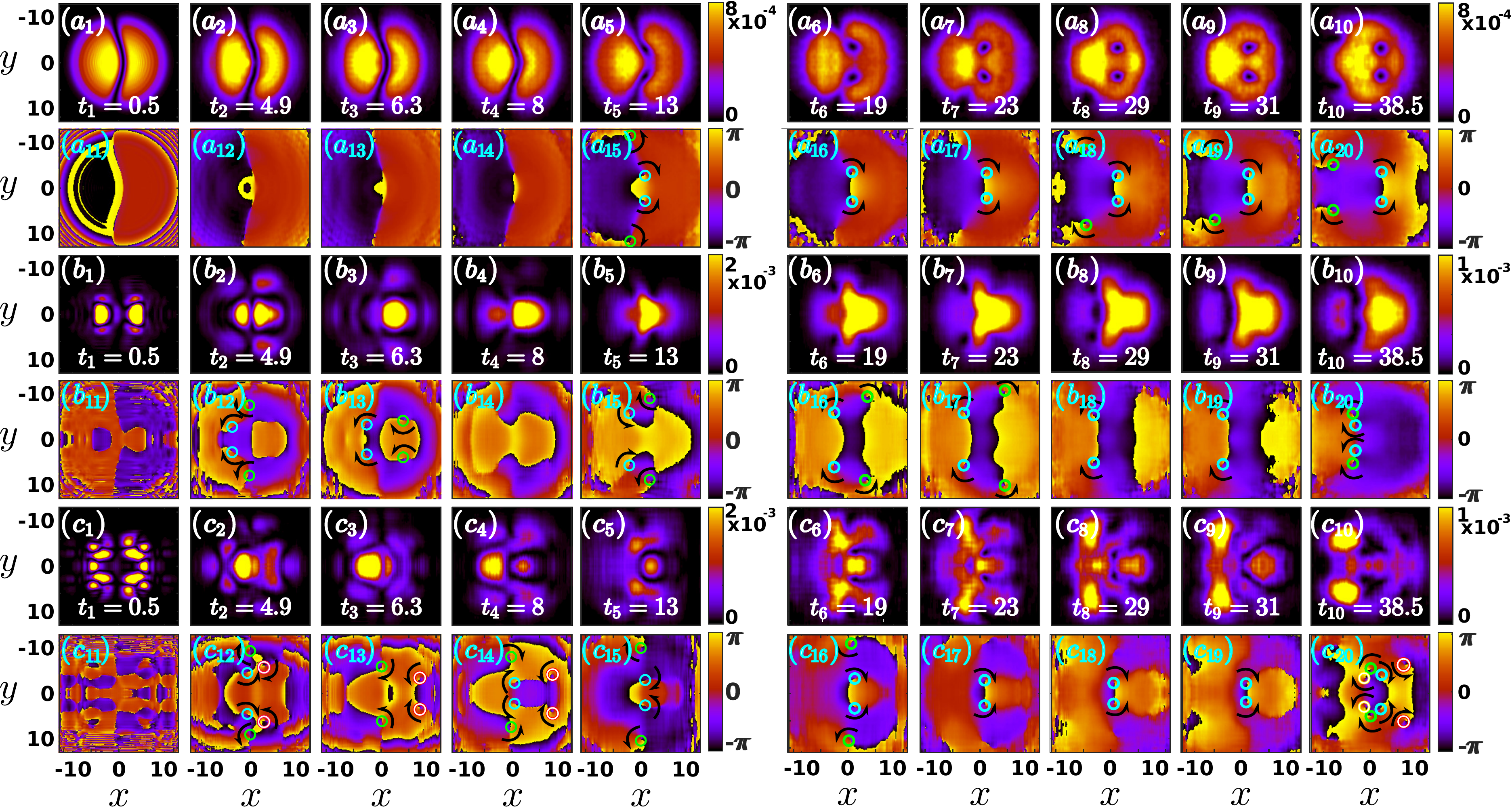}
\caption{(Color online) $|\phi_i(\textbf{r};t)|^2$, $i=1,2,3$ accompanied by the corresponding 
$\arg{[\phi_i(\textbf{r};t)]}$ at 
different time instants during evolution (see legend).  
In particular, $(a_1)-(a_{10})$, $(b_1)-(b_{10})$, $(c_1)-(c_{10})$ represent the 
first $|\phi_1(\textbf{r},t)|^2$, the second $|\phi_2(\textbf{r},t)|^2$ and the third $|\phi_3(\textbf{r},t)|^2$ orbital 
density respectively.  
In the same way, $(a_{11})-(a_{20})$, $(b_{11})-(b_{20})$, $(c_{11})-(c_{20})$ depict the corresponding instant phases of the first 
$\arg[\phi_1(\textbf{r},t)]$, second $\arg[\phi_2(\textbf{r},t)]$ and third $\arg[\phi_3(\textbf{r},t)]$ orbital 
density. Solid circles indicate the location of vortices while arrows show their circulation. 
Other parameters used are the same as in Fig.~\ref{Fig:1}. }
\label{Fig:3}
\end{figure*}

Next, let us demonstrate how the fragmentation of the system and consequently the MB character of the dynamics 
can be revealed by performing in-situ single-shot measurements \cite{kaspar,Lode}.  
The single-shot simulation procedure relies on a sampling of the MB probability distribution being available via MCTDHB. 
Referring to a fixed time instant $t_{im}$ of the imaging, first we calculate the one-body density 
$\rho^{(1)}_N(\textbf{r};t_{im})$ of the system from the MB wavefunction 
$\ket{\Psi_N}\equiv \ket{\Psi(t_{im})}$.  
Then, a random position $\textbf{r}'_1$ is drawn obeying the constraint $\rho^{(1)}_N(\textbf{r}'_1;t_{im})>z$ where $z$ 
refers to a random number within the interval [0, max\{$\rho^{(1)}_N(\textbf{r};t_{im})\}$].  
Next, one particle is annihilated at position $\textbf{r}'_1$ and the one-body density $\rho^{(1)}_{N-1}(\textbf{r};t_{im})$ 
of the reduced $N-1$ body system is calculated from $\ket{\Psi_{N-1}}$ and a new random position $\textbf{r}'_2$ is drawn from 
$\rho^{(1)}_{N-1}(\textbf{r};t_{im})$. 
In total, the procedure is repeated for $N-1$ steps and the resulting distribution of 
positions ($\textbf{r}'_1$, $\textbf{r}'_2$,...,$\textbf{r}'_{N-1}$) is convoluted with a point spread function 
to obtain a single shot $\mathcal{A}(\tilde{\textbf{r}})$ (for more details see \cite{kaspar,Lode}), where $\tilde{\textbf{r}}$ denote the 
spatial coordinates within the image.  
The employed spread function, here, consists of a Gaussian possessing a width $w\ll l_q$. 
To assess fragmentation from experimental single shot measurements we employ their variance 
for each time instant during the evolution of the BDS. 
The variance of a set of single shot measurements $\{\mathcal{A}_k(\tilde{\textbf{r}})\}_{k=1}^{N_{shots}}$ reads 
\begin{equation}
\mathcal{V}(t_{im})=\int d\tilde{\textbf{r}} \frac{1}{N_{shots}} \sum_{k=1}^{N_{shots}} [\mathcal{A}_k(\tilde{\textbf{r}};t_{im})-\bar{\mathcal{A}}(\tilde{\textbf{r}};t_{im})]^2,  
\end{equation}
where $\bar{\mathcal{A}}(\tilde{\textbf{r}};t_{im})=1/N_{shots}\sum_{k=1}^{N_{shots}} \mathcal{A}_k(\tilde{\textbf{r}};t_{im})$.  
Fig. \ref{Fig:2} ($f$) presents $\mathcal{V}(t)$ with $w=0.5$ and $N_{shots}=1000$ both at the MF and the MB level. 
As it can be seen within the MF approximation $\mathcal{V}(t)$ is approximately constant possessing negligible amplitude fluctuations. 
However when correlations are included $\mathcal{V}(t)$ exhibits an overall increasing tendency,  
resembling in this manner the fragmentation process, compare Figs. \ref{Fig:2} ($e$), ($f$). 
This similar behaviour of the fragmentation process and $\mathcal{V}(t)$ can be explained as follows.  
In a coherent condensate i.e. $n_1(t)=1$, $\mathcal{V}(t)$ is essentially constant during the evolution 
as all the atoms in the corresponding single shot measurement are drafted from the same SPF  
here $\varphi(t)$ [see also Eq. (\ref{Eq:MFansatz})].  
However, for a MB system where fragmentation is possible the corresponding MB state consists of a superposition 
of several configurations involving mutually orthonormal SPFs $\varphi_i(t)$, $i=1,...,M$ [see also Eq. (\ref{Eq:MCansatz})].  
Then, the variance of the single shots changes drastically from its MF counterpart because the atoms are picked from the above-mentioned superposition 
and therefore the distribution of the atoms in the cloud depends strongly on the position of the already imaged atoms.   
In addition, $\mathcal{V}(t)$ increases in time which can be attributed to the build up of higher-order superpositions 
in the course of the dynamics. 
We note that the above-described overall increasing behaviour of $\mathcal{V}(t)$ persists also for smaller samplings of single shot 
measurements namely $N_{shots}=200$, see Fig. \ref{Fig:2} ($f$).  
To showcase the robustness of the behaviour of $\mathcal{V}(t)$ with respect to the experimental resolution we present 
$\mathcal{V}(t)$ for $w=1$ and $N_{shots}=1000$, see Fig. \ref{Fig:2} ($f$), where the same increasing tendency as before is observed.  

Having established that the correlated character of the dynamics can be inferred from $\mathcal{V}(t)$, 
we next investigate whether the filling of the vortex cores can be observed by averaging several single shot images. 
We remark here that due to the diluteness of the considered bosonic gas $N=100$ the observation of the one-body density 
dynamics via a single shot image is not possible.  
To properly capture this dynamics via a single shot image a much higher particle number, e.g. $N\sim10^4$ is required. 
However in such a case the inclusion of more than two SPFs is computationally prohibitive and therefore numerical convergence 
on the MB level can not be ensured. 
On the other hand, the density obtained by averaging various single shot images suffers from unavoidable noise sources in the 
experiment, the most important of which is the optical resolution.   
The latter can give rise to an apparent filling of the vortex core even if the dynamics is of pure MF character. 
In the following we demonstrate how one can use the resolution of the image, namely the width $w$ of the point spread 
function, to resolve this issue in the averaged image. 
Figs. \ref{Fig:2} ($g$), ($h$), ($i$) show within the MF approximation the obtained average over $N_{shots}=1000$ images 
$\bar{\mathcal{A}}(\tilde{\textbf{r}};t_{im}=31)$ 
for increasing resolution i.e. decreasing $w$. 
As it is evident for $w=1 \sim \xi$ the core vortices also within the MF approximation possess a filled core, while for 
$w<\xi$ they are fully dipped thus recovering the well-known MF 
prediction. 
However within the MB approach the corresponding averaged images, see Figs. \ref{Fig:2} ($j$), ($k$), ($l$), exhibit filled 
core vortices for all considered resolutions.  

Concluding, in order to observe accurately the structures building upon the one-body density and discriminate the quantum 
features when employing an averaging of several 
single shot images one should use high resolution detectors. 
The latter can be accomplished by employing contemporary experimental techniques e.g. a quantum gas microscope 
\cite{qmicroscope1,qmicroscope2}. 
Finally, having at hand the averaged single shot images, one can directly measure the vortex dipole position 
both at and beyond the MF approximation. 
A shift on the vortex dipole location between the two approaches is observed, see for instance the dashed cyan lines at 
$x=1.26$ and $x=1.86$ in Figs. \ref{Fig:2} $(g)$ and $(j)$ respectively. 
Most importantly, the vortex dipole shift on the MB level when compared to the MF approach is robust independently of the 
imaging resolution, see also Figs. \ref{Fig:2} ($i$) and ($l$). 

\begin{figure*}[ht]
\includegraphics[width=1.0\textwidth]{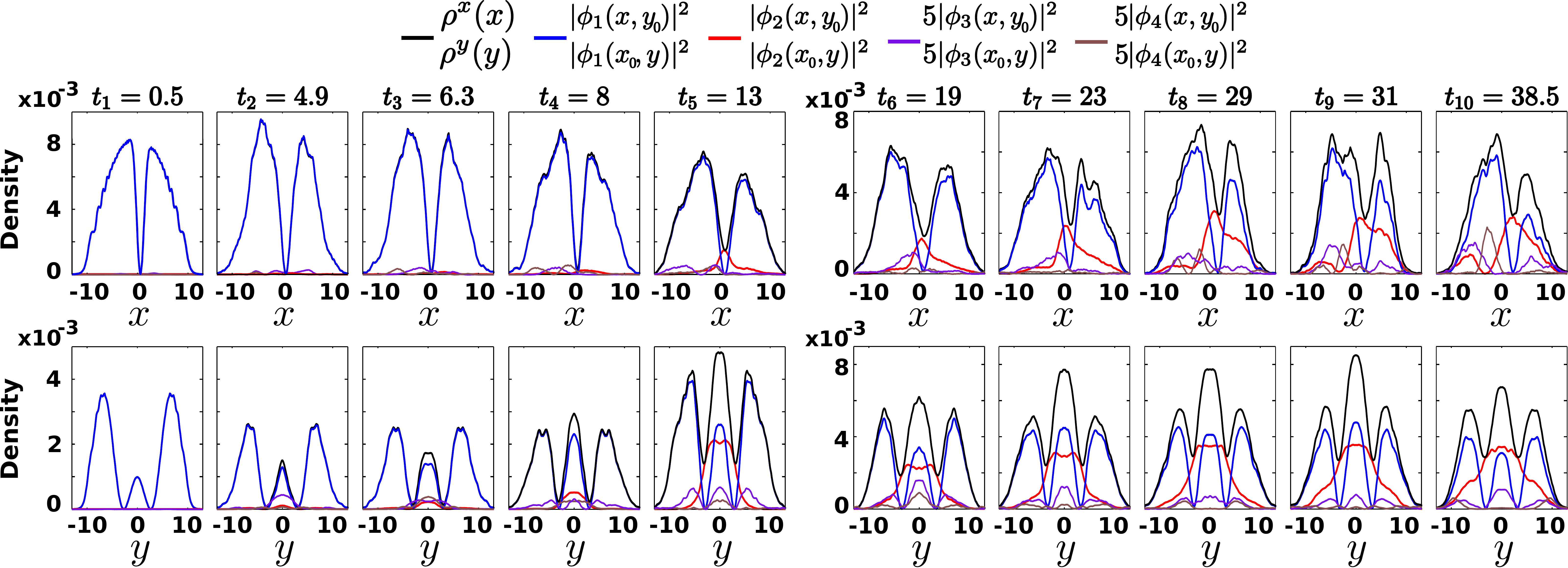}
\caption{(Color online): Density profiles corresponding to the same time instants depicted in Figs.~\ref{Fig:2}-\ref{Fig:3} 
respectively. In all panels shown are snapshots of the densities $\rho^{(1)}(\textbf{r};t)$, and $|\phi_i(\textbf{r};t)|^2$, 
with $i=1,2,3,4$ along the longitudinal, for fixed $y=y_0$, and transverse directions, 
for fixed $x=x_0$. 
Note that the density profiles of the third and 
fourth natural orbital are magnified by a factor of five to provide a better visibility of the structure that build upon them. 
In all cases the reference point $(x_0,y_0)$ is chosen around the region of maximum curvature of the initially embedded BDS, 
that corresponds to the location of the formation of the core vortex pair at later evolution times. 
Other parameters used are the same as in Fig.~\ref{Fig:1}.} \label{Fig:5}
\end{figure*}

\subsection{Orbital analysis}
Detailing the dynamics at the quantum level, we turn to the examination of the BDS snaking, its consequent 
decay, and the spontaneous vortex state nucleation in terms of orbitals. 
Note here that in order to compare the MF findings with the 
MB correlated approach, we used four natural orbitals. However, as already discussed above since only the first 
two orbitals are significantly occupied [see Fig.~\ref{Fig:2} $(d)$] we show in Fig.~\ref{Fig:3} 
representative time instants during propagation of the densities, $|\phi_i(\textbf{r};t)|^2$ with $i=1,2,3$, up to the third 
natural orbital. For completeness in the profiles depicted in Fig.~\ref{Fig:5} we also show the fourth natural orbital, as 
well as the total density, $\rho^{(1)}(\textbf{r};t)$, for the aforementioned time instants. 
To obtain the profiles of these five densities we choose as a reference point, $\left(x_0, y_0\right)$, the location of
maximum curvature of the initially embedded BDS.  
We remark here, that the first natural orbital as the leading order contribution, predominantly captures the MF picture.     
This result can easily be verified just by inspecting  
Figs.~\ref{Fig:3} $(a_1)$-$(a_{10})$ and comparing them with the relevant ones shown in Fig.~\ref{Fig:1}. 
As it is observed, e.g. in Figs.~\ref{Fig:3} $(a_2)$, $(a_3)$ and $(a_{12})$, $(a_{13})$ and also in the relevant profiles of 
Fig.~\ref{Fig:5} at $t=t_2$ and $t=t_3$ respectively, the BDS deforms
in the vicinity of its core soon after the beginning of the 
dynamics. This deformation, in accordance with the MF case, is followed by the creation 
of the core and edge vortex dipoles discussed above, which are clearly visible in Fig.~\ref{Fig:3} $(a_5)$ and 
in the corresponding phase depicted by circles in Fig.~\ref{Fig:3} $(a_{15})$. 

In parallel to that at this early stage of the dynamics, also the other orbitals build up. 
In particular, for times up to $t=t_5$ in both the second (predominantly occupied) and the third (not 
significantly populated) orbital illustrated in Figs.~\ref{Fig:3} $(b_{1})$-$(b_{5})$ and $(c_{1})$-$(c_{5})$ respectively, 
open ring dark solitonic structures~\cite{anderson} as well as oblique dark soliton-like patterns~\cite{el,amo}, 
are spontaneously formed. These patterns are clearly visible in e.g. Fig.~\ref{Fig:3} $(b_{3})$ and its 
phase in Fig.~\ref{Fig:3} $(b_{13})$ where both such structures are present, and/or also in Figs.~\ref{Fig:3} $(c_{4})$-$(c_{5})$. 
Notice that at $t=t_5$ two vortex pairs in each of the three natural orbitals are clearly formed.
Importantly, the location of the formation of these vortex dipoles differs for the different orbitals. 
In particular, the core vortex pairs developed in the 
first and third orbital are aligned, in contrast to the vortex dipoles nucleated in the second orbital 
[see Figs.~\ref{Fig:3} $(a_{15})$, $(b_{15})$, and $(c_{15})$ and also the relevant profiles 
  depicted in Fig.~\ref{Fig:5}].
In turn, the second orbital develops an antidark structure in the location of the formation 
of the core vortex dipole of the first orbital, ``filling'' in this way the vortices of the leading order orbital and resulting in the filled core vortex
density structure advertised earlier and clearly observed in Fig.~\ref{Fig:5},
as well as earlier in Fig.~\ref{Fig:2} $(a)$. The observed
vortex-antidark multi-orbital waveform
has the antidark structure placed only slightly off-center with respect to each vortex core.
It is worth mentioning at this point, that in contrast to the 
core vortex dipoles of the total density of Fig.~\ref{Fig:2} $(a)$, 
the vortex pairs created in individual orbitals are ``fully'' 
dipped as can also be seen in the profiles depicted in Fig.~\ref{Fig:5}. We note that a similar mechanism with dark solitons 
in the first orbital creating an effective double well potential trapping in turn antidark like structures created in higher-
lying orbitals was also observed in~\cite{lgspp} but in the 1D case. This filling mechanism seems to be rather generic, being also observed e.g. between 
the second and the third natural orbital this time with the vortices of the former 
filled by density humps of the latter as illustrated in Figs.~\ref{Fig:3} $(b_{5})$, $(c_{5})$ and in the corresponding phases 
of Figs.~\ref{Fig:3} $(b_{15})$, $(c_{15})$. Finally, the fourth orbital possessing a negligible occupation develops 
antidark structures aligned with the vortex pair of the third orbital as can be seen in Fig.~\ref{Fig:5} at $t=t_4$.
\begin{figure*}[ht]
\includegraphics[width=1.0\textwidth]{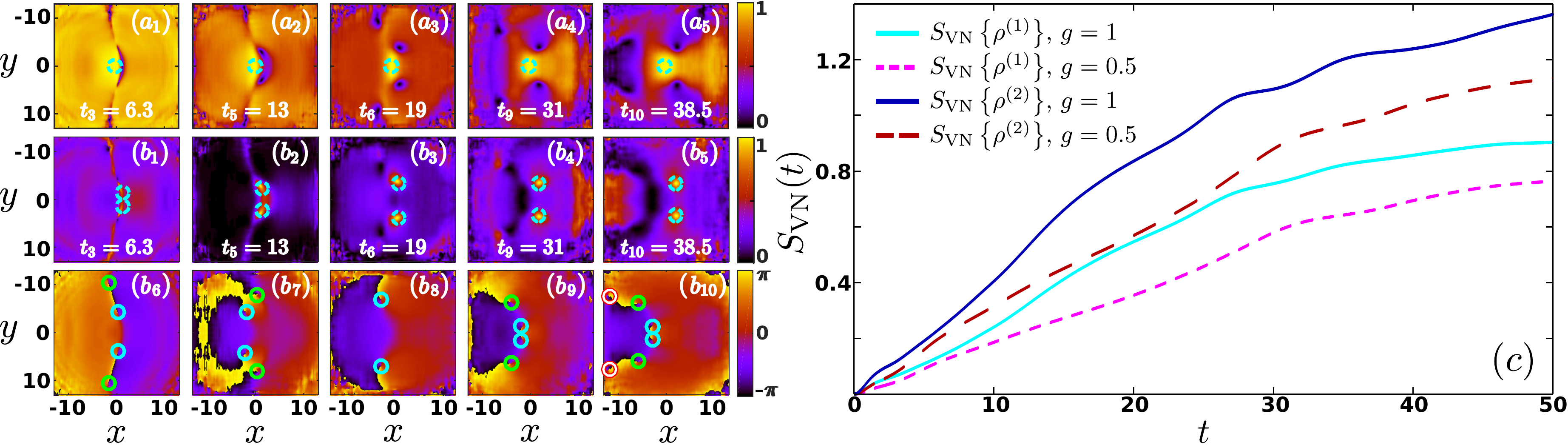}
\caption{(Color online) ($a_1$)-($a_5$) One-body coherence function, $g^{(1)}({\bf r}, {\bf r'};t)$, plotted at different time 
instants during the evolution (see legend), using as reference point
the center of the harmonic trap, ${\bf r'_1}=(0,0)$. 
($b_1$)-($b_5$) The same as the above but taking as a reference point, ${\bf r'_2}$, the location of 
the core vortex pair of the first natural orbital. 
The corresponding phases, $\arg{[g^{(1)}(\textbf{r}, {\bf r'_2};t)]}$, in 
this case are illustrated in ($b_6$)-($b_{10}$). In all cases dashed circles depict the reference points 
while the solid ones indicate the location of the vortices being visible in ($b_1$)-($b_5$).  
$(c)$ Evolution of the Von-Neumann entropy, $S_{VN}$, on the one- and two-body level for different interparticle repulsions 
(see legend). 
Other parameters used are the same as in Fig.~\ref{Fig:1}.} 
\label{Fig:6}
\end{figure*}

The aforementioned findings persist for larger propagation times, i.e. $t_6\leq t \leq t_{10}$, 
depicted in Fig.~\ref{Fig:3}. Namely, the vortex-antidark structure formed between the first and the second orbital 
respectively remains quite robust during evolution as is evident in the relevant profiles shown in Fig.~\ref{Fig:5}.
It is important to note here that as a result of the interaction between
the vortices and of the vortices with the background of their own,
as well as of other orbitals (e.g., the antidark structure), they 
gradually shift towards the positive $x-$direction. 
Interestingly enough, as time evolves and the second orbital becomes gradually more populated, this antidark structure 
overfills the core vortex dipole a result that is visible e.g. in Fig.~\ref{Fig:5} at $t=t_{10}$.
In the same time interval, the core vortex pair of the first orbital is supported 
by the core vortex dipoles of the third natural orbital, with the latter being aligned with the former at all times. 
Additionally, at these later time instants, a smearing effect of the edge vortex pair of the leading-order  
orbital is observed. Here, the edge vortex dipoles of the first orbital are partially filled by a density hump developed 
in the second orbital. This partial filling is also supported by humps created in the third orbital
as can be seen e.g. in Figs.~\ref{Fig:3} $(a_{8})$, $(b_{8})$, and $(c_{8})$ together with their corresponding phases shown
in Figs.~\ref{Fig:3} $(a_{18})$, $(b_{18})$, and $(c_{18})$.
Similarly, both vortex pairs of the second orbital remain filled with antidark-like structures created in the third orbital
and so on. 
Furthermore, oblique soliton patterns can also be observed to be
present in the late stages of propagation shown e.g. in Figs.~\ref{Fig:3} $(c_{9})$ and $(b_{9})$.
Such patterns do not persist but rather recombine from and split back
into vortex dipole pairs.
It is important to remark here that besides the vortices that support the leading order vortex dipoles, all the other
vortices belonging to the aforementioned cluster  
are never directly imprinted in the one-body density of the MB system~\cite{weiner}.
However, a careful inspection of the location of the formation of these hidden vorticity states reveals that these vortex 
dipoles are always created at locations not only shifted with respect to the leading order ones, but also
in regions where the lower-lying orbitals, which are predominantly occupied, developed the antidark entities.
As such, these vortex dipoles are immediately filled by both the lower and the higher-lying orbitals. 
The formation of density hump structures and vortex states holds also for the fourth orbital as can be seen in its magnified 
version depicted in Fig. ~\ref{Fig:5}. For instance,
at $t=t_{10}$, this orbital develops also a density hump that fills (but not significantly due to its population) the 
vortices created in the second natural orbital.

\subsection{Correlation analysis}
To investigate in more detail the localization mechanism observed in the orbital analysis 
during the BDS dynamics, we employ the normalized first order correlation function  
\begin{equation}
g^{(1)}({\bf r},{\bf r'};t)=\frac{\rho_{1}({\bf r}, {\bf r'};t)}{\sqrt{\rho_1({\bf r};t)\rho_1({\bf r'};t)}},
\end{equation}
which essentially measures the proximity of a MB state to a MF state for a given set of 
coordinates ${\bf r}$, ${\bf r'}$ and can be inferred via interference experiments~\cite{Hofferberth}. 
Note that $|g^{(1)}({\bf r}, {\bf r'} ) |^2$ is bounded, taking values within the interval $[0, 1]$.
A spatial region with $|g^{(1)}({\bf r}, {\bf r'} )|^2=0$ is referred to as
perfectly incoherent, while if $|g^{(1)}({\bf r}, {\bf r'} )|^2=1$, it is said to be fully coherent.  
In order to make an intuitive interpretation of this quantity, we use 
a fixed reference point. Figs. \ref{Fig:6} ($a_1$)-($a_5$) present $|g^{(1)}({\bf r},{\bf r'};t)|^2$ 
at selected time instants during the evolution for 
the reference point ${\bf r'_1}=(0,0)$, namely near the core of the initially embedded BDS.   
At the initial time instants, see Figs. \ref{Fig:6} ($a_1$) and ($a_2$), we observe the 
appearance of a smooth incoherent curved region which corresponds to the 
location of the BDS and it is the prominent feature of the first orbital 
[see also Figs. \ref{Fig:3} ($a_1$)-($a_5$)]. 
As time evolves, see Figs. \ref{Fig:6} ($a_3$)-($a_5$), the aforementioned incoherent 
region breaks into two fully incoherent pairs located in the left and right vicinity of the origin with respect to the $x$-
axis.  
These pairs correspond to the core and edge vortex pairs of the 
first orbital respectively, see also Fig. \ref{Fig:3}.  
We remark here that the same overall dynamics in terms of $|g^{(1)}({\bf r},{\bf r'};t)|^2$ is observed for all reference 
points ${\bf r'}$ located in the vicinity of
the BDS soliton (results not shown here for brevity).
However, differences in the observed dynamics occur upon considering as a reference point the location of the core 
vortex pair (${\bf r'}={\bf r'_2}$) of the first orbital, see Figs. \ref{Fig:6} ($b_1$)-($b_5$).  
We observe that at and in the proximity of the first orbital's core 
vortex pairs $|g^{(1)}({\bf r},{\bf r'=r'_2};t)|^2 \approx 1$ while away from these regions of vorticity,   
namely $|{\bf r-r'_2}|^2 \gg 0$,  
$|g^{(1)}({\bf r},{\bf r'=r'_2};t)|^2 \ll 1$ or even tends to zero throughout the evolution.
The emergence of spatially localized one-body correlations in the vicinity of
${\bf r'_2}$ manifested by the decay of the coherence function, $|g^{(1)}({\bf r}, {\bf r'_2} )|^2\to 0$,
as $|{\bf r-r'_2}|^2 \gg 0$ constitutes a key observation for the identification of 
localized structures namely the antidark states which appear in the second orbital 
[e.g. see Figs. \ref{Fig:3} ($b_6$)-($b_{10}$) and Figs. \ref{Fig:5} at $t_2$-$t_5$].    
More importantly, as time evolves the above-mentioned coherent regions are more prominent and expand around the core vortex 
pair of the first orbital. 
This expansion suggests that the localized antidark structures, as time evolves, can not be supported/trapped indefinitely by the first 
orbital and as a consequence diffuse within the cloud, see also the orbital structure in Fig.~\ref{Fig:5}.   
Besides the existence of the above described coherent regions we observe also the 
appearance of fully incoherent regions, especially for propagation times that the 
fragmentation manifests itself, see Figs.~\ref{Fig:6} ($b_1$)-($b_3$). 
A careful inspection of the location of these incoherent regions reveals that they reside in the region where the second 
orbital exhibits vortex pairs, see also Figs. \ref{Fig:3} ($b_6$)-($b_{10}$), which are not visible in the total density. 
To further elaborate on the existence of these vortex pairs we also show in Figs. \ref{Fig:6} ($b_6$)-($b_{10}$) the 
corresponding phases of $g^{(1)}({\bf r},{\bf r'_2};t)$. 
Note that the higher orbital structures possessing a small contribution 
compared to the second orbital, see Fig. \ref{Fig:2} 
($d$), are also imprinted in $|g^{(1)}({\bf r},{\bf r'_2};t)|^2$ 
to a minor extent as regions with lower coherence namely $|g^{(1)}({\bf r},{\bf r'_2};t)|^2\approx 0.5$.   
Concluding we can infer that by monitoring the coherence using as a fixed 
reference point the core vortex pairs of the first 
orbital both the localized antidark structures as well as 
the vortex pairs building upon the second orbital are visible.  

To further elaborate on the emergence of correlations, during the BDS dynamics, 
on both the one- and two-body level we employ the Von-Neumann entropy 
of the one- and two-body reduced density matrix respectively~\cite{Zozulya1,Zozulya2,Liu1,Liu2}. 
The Von-Neumann entropy on the $b$-body level reads
\begin{eqnarray}
S_{VN}[\rho^{(b)}(t)] &=& -\rm{Tr} (\rho^{(b)}(t) \log[\rho^{(b)}(t)] \nonumber \\ 
&=&-\sum_{i=1}^{\mathcal{M}^{(b)}} n_i^{(b)}(t) \log[ n_i^{(b)}(t)], 
\label{VN}
\end{eqnarray}
where $\rho^{(b)}(t)$ refers to the $b$-body reduced density matrix with eigenvalues $n_i^{(b)}(t)$ \cite{Sakmann} and 
$\mathcal{M}^{(b)}$ denotes the dimensionality of the b-body Hilbert space. Note that within our MB ansatz 
[see Eq. (\ref{Eq:MCansatz})] $\mathcal{M}^{(b)}$ corresponds to the truncated b-body Hilbert space spanned by the $M$ orbitals 
namely, $\mathcal{M}^{(b)}=\binom{b+M-1}{M-1}$. 
According to Eq. (\ref{VN}), the Von-Neumann entropy takes values within the range [$0$, $\log \mathcal{M}^{(b)}$]. 
The case of $S_{VN}=0$ refers to a pure $b$-body density matrix, e.g. $\rho^{(1)}(\textbf{r},
\textbf{r}^{'};t)=N\phi_1(\textbf{r};t)\phi_1^*(\textbf{r}^{'};t)$, 
which further implies the absence of correlations in the system. 
In that light when $S_{VN}\neq 0$ deviations from the MF approximation take place in the system.
However, when $S_{VN}=\log \mathcal{M}^{(b)}$ the $b$-body density is extremely mixed, 
e.g. $\rho^{(1)}(\textbf{r},\textbf{r}^{'};t)=\sum_{i=1}^{M}\frac{N}{M}\phi_i(\textbf{r};t)\phi_i^*(\textbf{r}^{'};t)$, and the 
corresponding correlations on the $b$-body level between the respective subsystems are maximized. 

Focusing on the presence of one- and two-body correlations, $S_{VN}$ is bound to take values within the intervals 
$[0, 1.4]$ and $[0, 2.3]$ respectively. Fig.~\ref{Fig:6} $(c)$
illustrates $S_{VN}\lbrace\rho^{(1)}(t)\rbrace$ and $S_{VN}\lbrace\rho^{(2)}(t)\rbrace$ for interparticle repulsions $g=0.5,1$. 
A monotonic increase of $S_{VN}$ is observed both on the one- and two-body level, showcasing the degree of mixedness 
of the BDS state.
Additionally, $S_{VN}\lbrace\rho^{(1)}(t)\rbrace< S_{VN}\lbrace\rho^{(2)}(t)\rbrace$ holds, since the available and 
significantly occupied number states are increased on the two-body level. The above observations suggest that also higher than 
one-body correlations participate in the BDS dynamics, in sharp contrast to the MF approximation where no correlations are included.
Note also that both $S_{VN}\lbrace\rho^{(1)}(t)\rbrace$ and $S_{VN}\lbrace\rho^{(2)}(t)\rbrace$, for the evolution times 
considered herein, do not reach their permitted maximum values and therefore the MB state is not maximally mixed in either 
the one- nor the two-body level. 
Finally, $S_{VN}$ shows the same overall behaviour for smaller interactions 
as can be seen in Fig.~\ref{Fig:6} $(c)$, but it is significantly lower as compared to stronger interactions. Namely
weaker interactions give rise to a lower degree of correlations and vice versa. 

\section{Conclusions}

In the present work the dynamical deformation of BDSs when exposed to quantum fluctuations 
has been investigated. In particular, upon considering a harmonically confined repulsively interacting 2D BEC,
we systematically explored the BDS decay and the resulting vortex nucleation
both in the MF limit where a single orbital dictates the dynamics, and within a MB correlated multi-orbital approach
namely MCTDHB.
It is found that both approaches show a qualitatively good agreement in capturing the decay of the BDS.
However, significant deviations between the two occur during the vortex nucleation process.
During this stage fragmentation becomes significant in the correlated approach,
and is found to be enhanced upon increasing the strength of the interparticle repulsion, resulting in the formation  
of vortex dipoles (two in our particular case). One of these dipoles is created at the core and the other at the edges of the initially embedded BDS.
These dipoles bear two characteristics that designate their multi-orbital nature when compared to their MF counter-parts. 
They are found to possess filled cores (rather than being fully dipped
as in the MF case) and are also significantly shifted with respect to the MF vortex pairs.
The former smearing effect that constitutes one of our central findings owes its occurrence 
to the emergence of an antidark structure that dynamically develops in the next-to-leading order orbital,
effectively filling in this way the depleted regions of the leading order one. 
The quantum nature of these states can be experimentally detected by measuring the variance of single shot 
images and can be directly observed by averaging the latter using a high resolution optical apparatus.  
This filling mechanism is a rather generic feature,
being also observed in 1D settings where the role of vortices is played by quantum dark solitons.
More importantly a hierarchy between the natural orbitals can be drawn. It is observed that when in an orbital
vortex dipoles nucleate, its subsequent (higher) orbital develops in the location of the formation of these vortex 
states antidark structures. Since the location of the vortices nucleated in higher-lying orbitals differs from orbital to 
orbital, the locations of the antidark solitons formed also differ. A complex
subsequent motion ensues as a result of the vortex-vortex, vortex-background
density (both of these within the same orbital)
and inter-orbital interaction. 
To gain further insight regarding the antidark and vortex structures created in the higher-lying orbitals, we also monitored 
the one-body coherence function. By this inspection we were able to show that localized one-body correlations indicate the 
presence of the antidark structures, while incoherent regions correspond to the location of the higher-lying vortex dipoles.
Finally, the mixedness of the MB state, and as a consequence the presence of not only one- but also two-body correlations
were identified by measuring the Von-Neumann entropy, revealing a monotonic growth of correlations verifying in this 
way the observed deviations from the MF approximation.        

There are numerous interesting potential extensions of the present
work. As a first generalization one could further add in the current setting
an external repulsive potential barrier. Since it is a well-known result that 
dark solitons, at least within the MF level, can be stabilized under this external trapping~\cite{ma},
it would be particularly interesting to examine the BDS dynamics  
under such confinement conditions both at and more importantly beyond the MF approximation. In this setting, it will be interesting to examine whether
(at the MF level) such a ``defect'' potential could render the BDS
a stationary configuration and, if so, the corresponding stability properties.
Comparing these features with their MB counterparts would be of
interest in its own right as a future direction.
Furthermore, and also at the level of single component solitary waves, examining
the fate of excitations such as (the unstable) ring dark solitons~\cite{rings},
or the result of dragging an obstacle through a condensate to produce
vortical patterns~\cite{neely} might offer further insights on
the dynamics of vortices (and vortex-antidark entities) beyond
the MF limit. 
Another relevant generalization involves the 
case of 2D mixtures. In the latter setting, it is well-known
at the MF level that vortex-bright solitons (as well as pairs
thereof) exist as robust 
configurations~\cite{law,pola}. 
It would be particularly interesting to examine how the filling mechanism analyzed here for vortices, 
is altered by the presence of the bright soliton component and vice versa. 
\begin{figure*}[ht]
\includegraphics[width=1.0\textwidth]{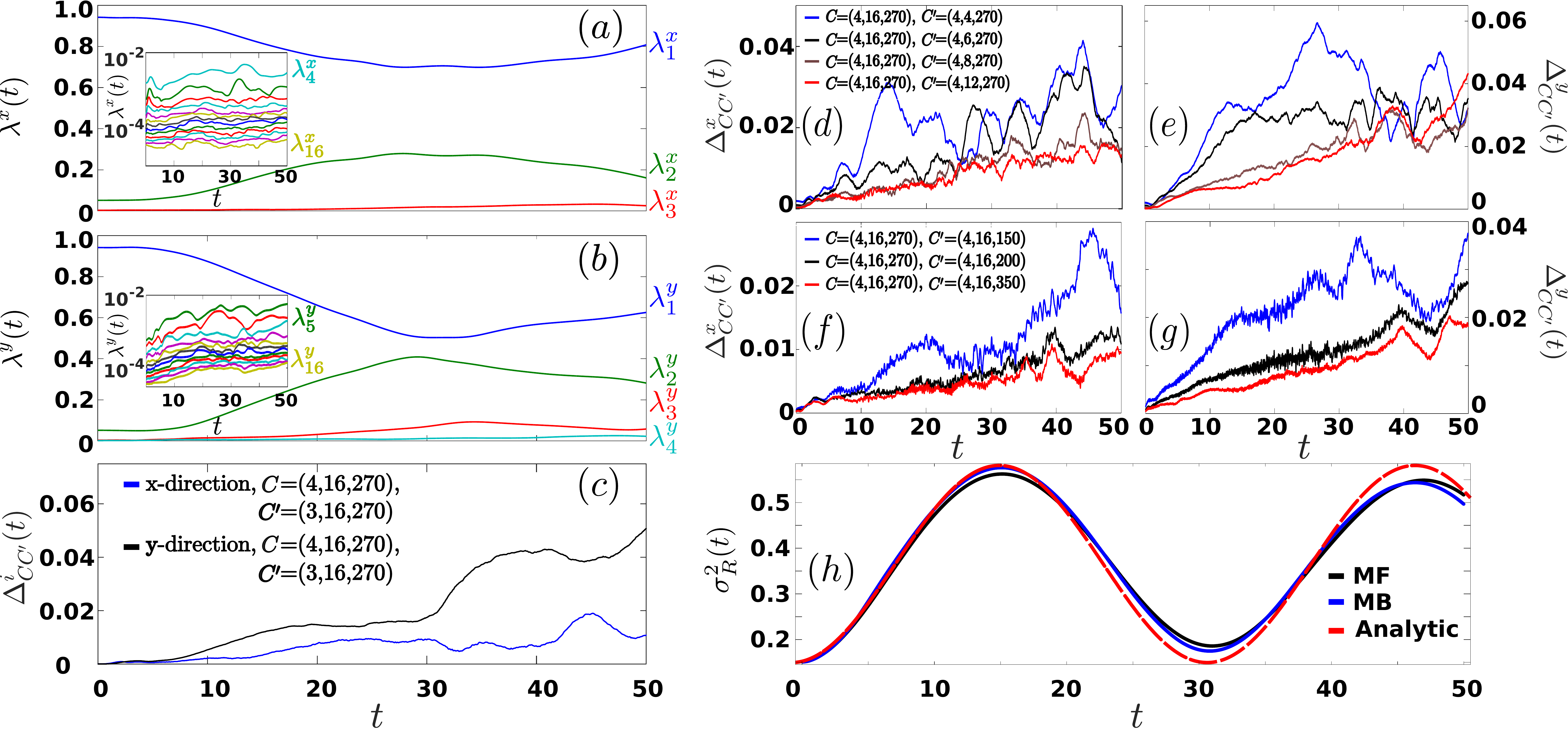}
\caption{(Color online)  
($a$) [($b$)] Time evolution of the first three [four] populations $\lambda_i^x(t)$ [$\lambda_i^y(t)$], $i=1,2,3$ 
of the reduced single particle density operator within the x [y] direction. 
The corresponding insets show the evolution of 
the $\lambda_{4}^x(t)$ to $\lambda_{16}^x(t)$ and 
$\lambda_{5}^y(t)$ to $\lambda_{16}^y(t)$ respectively. ($c$) Evolution of the spatially integrated differences 
$\Delta_{CC'}^x$ and $\Delta_{CC'}^y$ using $M=3$ and $M=4$ 2D SPFs. ($d$) [($e$)] $\Delta_{CC'}^x(t)$ [$\Delta_{CC'}^y(t)$] 
upon varying the number $m$ of the 1D SPFs (see legend). 
($f$) [($g$)] $\Delta_{CC'}^x(t)$ [$\Delta_{CC'}^y(t)$] between different number $M_p$ of grid sizes (see legend). 
($h$) Position variance $\sigma^2_R(t)$ during the BDS dynamics within the MB approach ($M=4$), the MF approximation  
and the analytical calculation (see legend). } 
\label{Fig:7}
\end{figure*}
\appendix

\section{Further details on the computational approach and convergence} \label{sec:numerics1}

In the present Appendix we outline some further features of our computational method (ML-MCTDHB) and elaborate on the 
convergence of our results.  

Within (ML-)MCTDHB the total MB wavefunction is expanded with respect to a time-dependent (t-d) variationally optimized 
MB basis. The latter allows us to span more efficiently the relevant, for the system under consideration, subspace 
of the Hilbert space at each time instant with a reduced number of basis states when compared to expansions relying on 
a time-independent basis.  
In particular, the MB wavefunction of $N$ bosons is expressed by a linear combination of t-d permanents $\ket{\vec n}_t$ with 
t-d coefficients $A_{\vec n}(t)$ 
\begin{equation}
 \ket{\Psi(t)}=\sum_{\vec n | \sum_i n_i = N} A_{\vec n}(t) \ket{\vec n}_t, 
\end{equation}
where the vector $\vec n=(n_1,n_2,...,n_M)$ and $n_i$ refers to the occupation of the $i$-th out of $M$ variationally 
optimized t-d 2D SPF $\ket{\varphi_i(t)}$.  
The summation is performed over all $N$-body permanents, i.e. all $n_i$'s such that they sum up to $N$.  

However, in the case of multi-dimensional systems excitations may not be isotropically spread along different spatial 
directions. Therefore, in order to have a more efficient treatment of the out-of-equilibrium dynamics it is more convenient to 
treat the induced excitations on the different spatial directions separately. 
The latter can be achieved within ML-MCTDHB \cite{moulos} by expanding each 2D t-d orbital $\ket{\varphi_i(t)}$, $i=1,2,...,M$ 
on two basis sets consisting  
of $m_x$, $m_y$ 1D t-d SPFs $\ket{\tilde \varphi_i^x(t)}$ and $\ket{\tilde \varphi_i^y(t)}$ respectively. 
Then, the corresponding SPF expansion reads     
\begin{equation}
\begin{split}
 \ket{\varphi_i(t)}=\sum_{j=1}^{m_x} \sum_{k=1}^{m_y} C_{i;jk}(t) \ket{\tilde \varphi_j^x(t)} \otimes \ket{\tilde 
 \varphi_k^y(t)}, 
\end{split}
\end{equation}
where $C_{i;jk}(t)$ refer to the corresponding t-d weights.  
Note here that in the present work we use the same number of 1D t-d SPFs in both directions, i.e. $m_x=m_y=m$.   
Finally, the 1D t-d SPFs $\ket{\tilde \varphi_i^{x,y}}$ are expanded with respect to a time-independent basis 
$\{\chi_{M_p}^{x,y}\}$.    
The latter basis is represented here by a one-dimensional sine discrete variable representation (DVR) grid consisting of 270 
grid points for each dimension. We remark here that our approach reduces to the usual MCTDHB 2D implementation if we supply as 
many 1D SPFs as the number of the used grid points i.e. $m_i=M_p$, while 
it is equivalent to the 2D GPE in case that we use only one 2D SPF $\ket{\varphi(t)}$.   

Next, let us comment on the convergence of our results upon varying the numerical configuration space $C=(M;m;M_p)$. 
We note here that all MB results presented in the main text rely on the configuration $C=(4;16;270)$. 
To infer that in the SPF expansion of Eq. ($A2$) the used number of the 1D t-d SPFs is sufficient, we examine the populations 
of the corresponding eigenvalues 
$\lambda_i^x(t)$, $\lambda_i^y(t)$, $i=1,...,m$ of the reduced density operator of a single boson within each direction.  
Figs. \ref{Fig:7} ($a$), ($b$) show the aforementioned populations during the dynamics for both directions. 
We observe that, within the $x$ ($y$) direction, the first three (four) SPFs are mainly populated and the remaining possess 
smaller amplitudes. In particular, the contributions of the last five eigenvalues i.e. $\lambda_{11}^{q}(t)-\lambda_{15}^{q}(t)$, 
q=x,y are negligible as they possess values below $10^{-4}$, see the insets of Figs. \ref{Fig:7} ($a$) and ($b$).  
To judge whether our calculations can be regarded as numerically converged, we 
demonstrate that the expectation value of the observables of interest become to a certain degree 
insensitive when increasing the number of basis states. In order to quantify the degree of convergence of the one-body density 
in each direction we employ the spatially integrated difference
\begin{equation}
 \Delta_{CC'}^{q}(t)=\frac{\int dq|\rho_{C}^{q}(t)-\rho_{C'}^{q}(t)|}{\int dq\rho_{C}^{q}(t)}, 
\end{equation}
where $\rho_C^{q}(t)$ [$\rho_{C'}^{q}(t)$] refers to the spatially integrated one-body density along the $q=x,y$ direction 
e.g. within the $x$-direction $\rho^x(t)=\int dy \rho^{(1)}(x,y;t)$. 
The calculations are performed within the configurations $C=(M;m;M_p)$ and $C'=(M';m';M'_p)$.  
Fig. \ref{Fig:7} ($c$) presents $\Delta_{CC'}^{q}(t)$ for both spatial directions within the numerical  
configurations $C=(4;16;270)$ and $C'=(3;16;270)$, i.e. increasing the number of the t-d 2D SPFs.  
As it can be seen $\Delta_{CC'}^{q}(t)$ testifies negligible deviations in both directions. 
In particular within the $x$-direction $\max[\Delta_{CC'}^x(t)]=1.8\%$ while in the $y$-direction, which is more prone to 
excitations, $\max[\Delta_{CC'}^y(t)]=6\%$ for long evolution times. 
The same observations hold for an increasing number of the t-d 1D SPFs $\ket{\tilde \phi_i^{x,y}(t)}$, 
see Figs. \ref{Fig:7} ($d$), ($e$). 
Indeed, $\Delta_{CC'}^{q}(t)$ shows a progressive convergence of $\rho_{C}^{q}(t)$ upon incrementing $m$.     
For instance, $\max[\Delta_{CC'}^x(t)]=1.8\%$ and $\max[\Delta_{CC'}^y(t)]=3\%$ between the configurations 
$C=(4;16;270)$ and $C'=(4;12;270)$. 
Finally, we examine the convergence of our results for different grid sizes, namely upon varying $M_p$.  
As shown in Figs. \ref{Fig:7} ($f$), ($g$) $\Delta_{CC'}^{q}(t)$ becomes fairly small for increasing $M_p$, 
e.g. $\max[\Delta_{CC'}^x(t)]=0.8\%$ and $\max[\Delta_{CC'}^y(t)]=1.5\%$ for 
$C=(4;16;270)$ and $C'=(4;16;350)$. 

To further elaborate on the convergence of our simulations we show the behaviour of the center of mass (CM) 
variance calculated both analytically (see below) and numerically. 
The harmonic oscillator potential allows for the separation of the CM, $R_q=\frac{1}
{N}\sum_iq_i$, and the relative coordinates $r_q=q_{i+1}-q_{i}$ where $q=x,y$.  
Then, the $N$-body interacting problem can be reduced to an interacting $N-1$-body problem in the relative 
coordinates, and a non-interacting one for the CM coordinate. 
However, our calculations within ML-MCTDHB have been performed in the lab frame and as a consequence do not 
utilize the aforementioned separation of variables. 
Note that both the ML-MCTDHB as well as the MF ans{\"a}tze do not trivially respect the separation between the CM and relative 
frame~\cite{Cosme}.  
Despite the above, as we shall show below our results can capture the decoupling of the CM motion for the entire $N$-body bosonic cloud.  
To judge about relative deviations of the ML-MCTDHB propagation with the full Schr{\"o}dinger equation (and consequently about 
convergence) we compare the ML-MCTDHB obtained evolution of the CM coordinate to the analytical one. 
The second moment of the CM position (position variance) reads 
\begin{equation} 
\begin{split}
\sigma_R^2(t)=&\sigma^2_{R_x}(t)-\sigma^2_{R_y}(t)\\&=\braket{R_x^2(t)}-\braket{R_x(t)}^2+\braket{R_y^2(t)}-\braket{R_y(t)}^2
\end{split}  
\end{equation} 
where $R_x$, $R_y$ denotes the mean position of the bosonic cloud in the $x$ and $y$ direction respectively.  

By using the Ehrenfest theorem on the CM Hamiltonian we obtain the exact evolution of the CM position 
variance 
\begin{equation}
\begin{split}
 \sigma_{R_q}^2(t)=&\left[\langle{R_q^2}\rangle(0)-[\langle{R_q}\rangle(0)]^2\right] \cos^2 \omega t \\ 
&+ \frac{1}{\omega^2} \left[\langle{P_q^2}\rangle(0)-[\langle{P_q}\rangle(0)]^2\right] \sin^2 \omega t \\ 
&+\frac{1}{2 \omega}\langle{ R_q P_{q'} + P_q R_{q'} }\rangle(0) \sin 2 \omega t \\ 
&-\frac{1}{\omega}\langle{ R_q}\rangle(0)\langle{ P_q}\rangle(0) \sin 2 \omega t. \label{Eq:B4} 
\end{split} 
\end{equation} 
$R_q$, $P_q$ with $q=x,y$ denote the spatial coordinate and the momentum operators within the q-direction   
acting on the CM degree of freedom. 
This latter expression offers the opportunity to directly measure the deviation between the MB approach, the
MF ansatz, and the analytical result. 
To expose this deviation we numerically 
calculate the position variance and compare with Eq. (\ref{Eq:B4}), see Fig.~\ref{Fig:7} ($h$).   
As it can be seen the MB result using four orbitals ($M=4$) follows the behaviour of the analytical 
result and therefore can be considered trustworthy.  
The observed maximum deviation at long propagation times $t>30$ is of the order of $6.5\%$. 
The MF result when compared to the correlated approach exhibits a slightly larger deviation compared to the analytics which at long evolution times becomes  
of the order of $8\%$. 
Summarizing, the above systematic investigations can guarantee the convergence of our results and as a consequence the 
robustness of the emerging structures in the beyond MF dynamics.

\section{Initialization of the beyond mean-field dynamics} \label{sec:numerics}

In the present section we briefly comment on our initial state preparation.  
To initialize the beyond MF dynamics we optimize the MF solution  
and embed it into the ML-MCTDHB ansatz, see Eq.~(\ref{Eq:MCansatz}). 

To obtain the MF state, the number of particles $N$ and the parametrized initial position $x(y)$ 
are kept fixed. The algorithm is initialized by assuming ansatz values for the background chemical potential $\mu^{(0)}$, 
while the inverse width is always set equal to $d=\frac{1}{\xi}$, where $\xi=\frac{1}{\sqrt{|\tilde{\phi}_0|^2 g}}$.
The structure of the algorithm proceeds as follows.
First we obtain the MF solution for the background density of the GPE for $\mu^{(0)}$ and $\mu^{(0)} +\delta \mu$
using a Newton type, fixed point algorithm. 
For the latter an underlying basis consisting of a $270\times 270$ numerical grid is used and a second order central finite 
difference method is employed in order to approximate the derivatives. 
Then, we calculate $N(\mu)=\int dx dy|\phi_{\mu}(x,y)|^2$, approximate $\frac{dN}{d\mu}$ and update the chemical potential. 
In turn, we iterate the above two steps until the particle number 
converges to $N$, thus obtaining the required 2D MF wavefunction, $\tilde \phi(x,y)$.
Having $\tilde \phi(x,y)$, the corresponding 2D one-body density matrix 
$\rho^{(1)}(x,x';y,y')=\tilde \phi^*(x',y')\tilde \phi(x,y)$ can be constructed. 
To obtain the corresponding reduced 1D eigenfunctions $\varphi^x_i(x)$, $i=1,\dots,m_x$  
we diagonalize the reduced 1D one-body density matrix 
namely $\rho^{(1)}_{x}(x,x')=\int dy~\rho^{(1)} (x,x';y,y)$. 
Given the two sets of reduced 1D eigenfunctions for the $x$ and $y$ axes we can express the 2D MF
wavefunction upon the basis spanned by the reduced 1D eigenfunctions 
as $\tilde \phi(x,y)=\sum_{j,k} C_{j,k} \tilde \varphi^x_j(x) \tilde \varphi^y_k(x)$. 

Finally, the solutions obtained by the above process are properly normalized and embedded as the first SPF of the ML-MCTDHB 
ansatz. The remaining initially unoccupied used SPFs are randomly-generated from a uniform distribution, i.e. $C_{i;j,k}(0)=random$ for $i \neq 1$,  
and are orthonormalized according to the Gram-Schmidt algorithm. 
To ensure that our results are independent of the above-mentioned randomization process we have used several different 
randomly generated states and we have obtained for each one the same evolution.  
In this way, the MB wavefunction is initialized in the state where all the particles reside in the corresponding first SPF, 
i.e. $A_{n_1=N}(0)=1$, $A_{n_1\neq N}(0)=0$ (see also text). 

\section*{Acknowledgements} 
The authors S.I.M. and P.S. gratefully acknowledge financial support by the Deutsche Forschungsgemeinschaft 
(DFG) in the framework of the
SFB 925 ``Light induced dynamics and control of correlated quantum
systems''. G.M.K and P.S. acknowledge support by the excellence cluster 
`` The Hamburg Center for Ultrafast Imaging: Structure, Dynamics and Control
of Matter at the Atomic Scale''. G.C.K. and P.S. gratefully acknowledge financial support by the DFG in the framework of the grant
SCHM 885/26-1. P.G.K. gratefully acknowledges the
support of NSF-DMS-1312856, NSF-PHY-1602994, the
Alexander von Humboldt Foundation, and the ERC under
FP7, Marie Curie Actions, People, International Research
Staff Exchange Scheme (IRSES-605096).

{}


\begin{thebibliography}{60}
\bibitem{mlmctdh} We note here, that the ML-MCTDHB method provides the variationally optimal total wavefunction for 
a given number of SPFs.
\end{thebibliography}
\end{document}